\documentclass[showpacs,twocolumn,aps,prl]{revtex4-2}
\usepackage{graphicx}
\usepackage{dcolumn}
\usepackage{bm}
\usepackage{amsmath}
\usepackage{amssymb}
\usepackage[usenames,dvipsnames]{xcolor}
\usepackage{color}
\usepackage[colorlinks,plainpages=false]{hyperref}

\begin{document}

\title{Strangelets at finite temperature in an equivparticle model}
\author{H.~M.~Chen,$^1$ C.J.Xia,$^2$ G.~X.~Peng,$^{1,3,4}$}

\affiliation{%
$^1$\mbox{School of Nuclear Science and Technology, University of Chinese Academy of Sciences, Beijing 100049, China}\\
$^2$\mbox{School of Information Science and Engineering, Ningbo Institute of Technology, Zhejiang University, Ningbo 315100, China}\\
$^3$\mbox{Theoretical Physics Center for Science Facilities, Institute of High Energy Physics, P.O. Box 918, Beijing 100049, China}\\
$^4$\mbox{Synergetic Innovation Center for Quantum Effects and Application, Hunan Normal University, Changsha 410081, China}  }

\begin{abstract}
The properties of strangelets at finite temperature are studied within the framework of an equivparticle model, where a new quark mass scaling and self-consistent thermodynamic treatment are adopted. The effects of finite volume and Coulomb energy are taken into account. Our results show that the temperature $T$, baryon number $A$ and perturbation interactions have strong influences on the properties of strangelets. It is found that the energy per baryon $M/A$ and charge-to-mass ratio $f_z$ decrease with baryon number $A$, while the mechanically stable radius $R$ and strangeness per baryon $f_S$ are increasing. For a strangelet with a fixed baryon number, we note that as temperature $T$ increases the quantites $M/A$, $R$, and $f_S$ are increasing while $f_z$ is decreasing. The effects of confinement and perturbative interactions are investigated as well by readjusting the corresponding parameters.
\end{abstract}


\maketitle

\section{Introduction}
\label{intro}

Since Bodmer studied the possible collapse of finite nuclei in 1971~\cite{Bodmer1971_PRD4-1601}, the properties of quark matter has begun to attract much attention, which is particularly the case after Witten proposed that strange quark matter (SQM) might be the true ground state of strong interaction in 1984~\cite{Witten1984_PRD30-272}. Compared with ordinary nuclear matter, the density of quark matter is much larger so that the Fermi-energy of $u$ and $d$ quarks can easily surpass the rest mass of $s$ quarks. Then it is energetically favorable to convert $u$ and $d$ quarks into $s$ quarks and form SQM. Farhi and Jaffe then investigated the properteis of SQM based on the MIT bag model and found that the energy per baryon can be lower than that of $^{56}$Fe (930 $\mathrm{MeV}$) in a large parameter space, i.e., SQM is absolutely stable \cite{Farhi_PhysRevD.30.2379}. If true, both strange stars and strangelets that made of SQM are expected to be observed in the Universe. Extensive studies on SQM were then carried out~\cite{Peng_PhysRevC.59.3452, Peng_PhysRevC.62.025801, Xu_PhysRevD.92.025025, Xia_PhysRevD.89.105027, XuPhysRevD.96.063016}.

Strange quark matter nuggets with baryon number $A$ less than $10^{7}$ are called strangelets \cite{Berger1987_PRC35-213}, or slets for short~\cite{Peng2006_PLB633-314}. There are mainly two methods to produce strangelets, i.e., by nature or by experiments. For example, it is expected that the collision of strange stars may eject strangelets, which could eventually reach earth~\cite{Banerjee1999_JPG25_L15, Banerjee2000_PRL85_1384--1387, Sandweiss_JPG_2003, Saito_PhysRevLett.65.2094, Madsen_PhysRevLett.90.121102, Madsen_PhysRevD.71.014026}. During the hadronization process of the early universe, strangelets may form and constitute dark matter~\cite{Merafina2020 PRD102-083015}. The collision of highly energetic cosmic rays with Earth's atmosphere may also produce strangelets~\cite{Madsen_PhysRevLett.90.121102}. Additionally, it is possible that strangelets are produced in heavy ion collision experiments~\cite{Rusek_PhysRevC.54.R15, Buren_JPG_1999, Appelquist_PhysRevLett.76.3907, WEINER2006_IJMPE15_37-70}.

At present, the MIT bag model has been widely used to study the properties of strangelets and obtained some important results. For example, Farhi and Jaffe found that the strangelets has a small amount of positive charge under weak equilibrium condition \cite{Farhi_PhysRevD.30.2379}. Greiner investigated the possibility to synthesize strangelets via heavy ion collisions \cite{Greiner_PhysRevLett.58.1825,Greiner_PhysRevD.44.3517}. Adopting the MIT bag model, Madsen considered the finite size effects of strangelets at vanishing temperatures~\cite{Madsen_PhysRevLett.70.391, Madsen_PhysRevD.47.5156, Madsen_PhysRevD.50.3328}, while Mustafa studied the stability of finite-size strangelets at finite temperatures \cite{Mustafa_PhysRevD.53.5136}. The effects of finite temperature on the stability and thermodynamical properties of strangelets were later investigated by Mustafa and Zhang~\cite{Mustafa_PhysRevD.53.5136, Zhang_PhysRevC.67.015202}. In addition, taking into account the contributions of electrostatic effects, Debye screening and nonzero surface tensions, Alford had found the critical value of surface tension to maintain the stability of strangelets~\cite{Alford_PhysRevD.73.114016}.

In addition to the MIT bag model, there are various effective models reflecting QCD characteristics, e.g., perturbation models \cite{Fraga2005}, quark-cluster model \cite{Shi2003,R.2010} and so on \cite{Plumari2013, Khadekar2011, Wen2013a, Isayev2013, Wang2010, Huang2004, Bao2008, Peng_PhysRevC.59.3452,Peng_PhysRevC.62.025801,Hou2015}. The properties of strangelets were thus investigated with those effective models, e.g., NJL model \cite{Yasui_PhysRevD.71.074009} and quasiparticle model \cite{Wen_PhysRevC.82.025809}.
The equivparticle model has also been used to study the properties of strangelets and many interesting conclusions have been drawn at $T=0$~\cite{Peng_PhysRevC.77.065807, Peng2010_IJMPE19_1837-1842}. In this paper, we apply the equivparticle model to study the properties of strangelets at finite temperature, where the effects of various quark interactions are examined by adopting different parameter sets after we consider the contributions of Coulomb interaction.

The paper is organized as follows. In Sec.~\ref{inconThermo}, we give the thermodynamic treatment of strangelets at finite temperatures in the framework of equivparticle model, where the contributions of Coulomb interaction are accounted for. In Sec.~\ref{sec:therm}, we consider the quark mass scaling and gluon mass scaling for SQM at finite temperatures, where both the confinement interaction and perturbation interaction are considered in the equivalent mass of quarks. The relationship between gluon mass and temperature is discussed as well. In Sec.~\ref{sec:SLT}, we present the numerical results on the properties of strangelets at finite temperature, where the energy per baryon, radius and charges of strangelets are discussed. Finally, a summary is given in Sec.~\ref{sec:sum}.

\section{Self-consistent thermodynamic treatment}
\label{inconThermo}

The most complex and controversial issue in equivparticle models is thermodynamic self-consistency, where various thermodynamic treatments were developed based on different considerations~\cite{PLB_CHAKRABARTY1989112,Benvenuto_PhysRevD.51.1989,Peng_PhysRevC.62.025801,Wen_PhysRevC.72.015204}. In this paper, we start from the free energy density $F$ and obtain the other thermodynamic quantities through self-consistent thermodynamic treatment, i.e., taking $F$ as the characteristic thermodynamic function. The free energy density is expressed in the same form as the free particle system, but the mass is replaced by the equivalent one which depends on temperature and density. The free energy density is given by
\begin{eqnarray}\label{1}
F &=& F(T,V,\{n_{i}\},\{m_{i}\})\nonumber\\
  &=& {}\Omega_{0}(T,V,\{\mu_{i}^{*}\},\{m_{i}\})+\sum_{i=u,d,s}\mu_{i}^{*}n_{i}.
\end{eqnarray}
where $T$ is the temperature, $n_{i}$ the particle number density,  $m_{i}$ the mass, and $\mu_i^*$ the effective chemical potential of particle type $i$. Note that $\Omega_{0}$ corresponds to the thermodynamic potential density of a system comprised of noninteracting particles, which should be viewed as an intermediate variable instead of the real one. The corresponding differential equation is then obtained with
\begin{eqnarray}\label{2}
\mathrm{d}F &=& \frac{\partial\Omega_{0}}{\partial T}\mathrm{d}T+\frac{\partial\Omega_{0}}{\partial V}\mathrm{d}V\nonumber\\
          & &{} +\sum_{i}\left(\frac{\partial\Omega_{0}}{\partial \mu_{i}^{*}}\mathrm{d}\mu_{i}^{*}+\mu_{i}^{*}\mathrm{d}n_{i}+n_{i}\mathrm{d}\mu_{i}^{*}\right)\nonumber\\
          & &{} +\sum_{i}\frac{\partial\Omega_{0}}{\partial m_{i}}\left(\sum_{j}\frac{\partial m_{i}}{\partial n_{j}}\mathrm{d}n_{j}+\frac{\partial m_{i}}{\partial T}\mathrm{d}T\right) \nonumber\\
          &=& \left[
      \frac{\partial \Omega_0}{\partial T}
      + \sum_i \frac{\partial \Omega_0}{\partial m_i} \frac{\partial m_i}{\partial T}
\right]\mbox{d}T
+ \frac{\partial \Omega_0}{\partial V} \mbox{d}V
\nonumber \\
&&
+
\sum_i \left[ \mu_i^* +  \sum_j\frac{\partial \Omega_0}{\partial m_j}\frac{\partial m_j}{\partial n_i}
\right] \mbox{d} n_i.
\end{eqnarray}
Comparing with the basic thermodynamic differential relation
\begin{eqnarray}\label{4}
\mathrm{d}F
&=&-S\mathrm{d}T+\left(-P-F+\sum_{i}\mu_{i}n_{i}\right)\frac{\mathrm{d}V}{V}\nonumber\\
& &+\sum_{i}\mu_{i}\mathrm{d}n_{i},
\end{eqnarray}
we can then obtain entropy density $S$, pressure $P$, and chemical potential $\mu_{i}$ with
\begin{eqnarray}
S &=& -\frac{\partial\Omega_{0}}{\partial T}-\sum_{i}\frac{\partial m_{i}}{\partial T}\frac{\partial\Omega_{0}}{\partial m_{i}}, \label{6} \\
P &=& -F+\sum_i \mu_{i} n_i - V\frac{\partial \Omega_0}{\partial V}, \label{7} \\
\mu_{i} &=& \mu_i^* +   \sum_j\frac{\partial \Omega_0}{\partial m_j}\frac{\partial m_j}{\partial n_i}. \label{8}
\end{eqnarray}
The particle number densities $n_{i}=n_i^+-n_i^-$ and the energy density $E$ are then fixed by
\begin{eqnarray}
n_{i}^{\pm} &=&  -\frac{\partial \Omega_{0}^{\pm}}{\partial \mu_{i}^{*}}, \label{5} \\
E &=& F+TS=\Omega_{0}+\sum_{i}\mu_{i}^{*}n_{i}+TS. \label{10}
\end{eqnarray}

Because the defined strangelets baryon number $A$ is less than $10^{7}$, its radius is usually smaller than the Compton wavelength of the electron, so that electrons are not confined within the strangelets. Then we can neglect electrons and the corresponding chemical potential is exactly zero if we minimize the mass of a strangelet. This property has two implications, i.e., the relaxation of the local charge neutrality condition and the contributions of Coulomb interaction in the weak-equilibrium condition for strangelets.
Consider the contribution of volume charge, the interaction energy between volume charge and surface charge, and the contribution of surface charge, we can obtain the Coulomb energy as follows
\begin{eqnarray}\label{13}
\bar{E}_\mathrm{C}
 &=&
 \frac{3\alpha_\mathrm{C} \bar{Q_\mathrm{v}}^2}{5 R}
 +
 e \phi_\mathrm{v}(R) \bar{Q_\mathrm{s}}
 +
 \frac{e \phi_\mathrm{s}(R) \bar{Q_\mathrm{s}}}{2}
\nonumber\\
 &=&
 \frac{\alpha_\mathrm{C}}{2 R}(\bar{Q}^2+\frac{1}{5}\bar{Q_\mathrm{v}}^2),
\end{eqnarray}
where $e \phi_\mathrm{v}(R)=\alpha_\mathrm{C} \bar{Q_\mathrm{v}}/R,\
 e \phi_\mathrm{s}(R)=\alpha_\mathrm{C} \bar{Q_\mathrm{s}}/R$. $\bar{E}_\mathrm{C}$ is the total energy of Coulomb, $\alpha_{C}\approx 1/137$ represents the fine structure constant. Then we have
 \begin{eqnarray}\label{14}
E_\mathrm{C}
 &=&
 \frac{3\bar{E}_\mathrm{C}}{4\pi R^3}
 =
 \frac{2}{15}\pi R^2 \alpha_\mathrm{C}  (5 Q^2+Q_\mathrm{v}^2),
\end{eqnarray}
where ${E}_\mathrm{C}=\bar{E}_\mathrm{C}/V$ is the Coulomb energy density and $Q_{V}$ represents the volume contribution of total charge density $Q$, namely $Q=\sum_{i}q_{i}n_{i}$, and $Q_{V}=\sum_{i}q_{i}n_{i,V}$. The charge of each particle $q_{u}=\frac{2}{3}$, $q_{d}=-\frac{1}{3}$, $q_{s}=-\frac{1}{3}$. The total energy density of a strangelet is then obtained with
\begin{eqnarray}\label{15}
E_\mathrm{total}= E +E_\mathrm{C},
\end{eqnarray}
where $E$ is fixed by Eq.~(\ref{10}) and $E_{c}$ by Eq.~(\ref{14})

For the pressure, entropy and chemical potential, the contribution of the Coulomb energy can be obtained with Eqs.~(\ref{6})-(\ref{8}), which are expressed in terms of $P_\mathrm{C}$, $S_\mathrm{C}$ and $\mu_{i,\mathrm{C}}$, respectively. According to the basic differential relation of thermodynamics, they satisfy the following differential expression
\begin{eqnarray}\label{20}
\mbox{d} E_\mathrm{C}   &=&\left(T S_\mathrm{C}  -P_\mathrm{C}-E_\mathrm{C}+ \sum_i \mu_{i,\mathrm{C}} n_i \right)\frac{\mbox{d}V}{V}
 \nonumber  \\
 & &{} +\sum_i \mu_{i,\mathrm{C}} \mbox{d} n_i+T \mbox{d}S_\mathrm{C}.
\end{eqnarray}
Meanwhile, the Coulomb energy density of a strangelet is a function of the quark number density and volume, i.e., $E_\mathrm{C}=E_\mathrm{C}(V, n_i)$. This indicates
\begin{eqnarray}\label{21}
  \mbox{d} E_\mathrm{C} &=& \sum_i\frac{\partial E_\mathrm{C}}{\partial n_i} d n_i
                   +\frac{\partial E_\mathrm{C}}{\partial V} \mbox{d} V.
\end{eqnarray}
Then we have
\begin{eqnarray}
  \mu_{i,\mathrm{C}} &=& \frac{\partial E_\mathrm{C}}{\partial n_i} = \frac{\partial E_\mathrm{C}}{\partial m_i}\left/ \frac{\partial n_i}{\partial m_i}=\frac{4}{3}\pi R^2 \alpha_\mathrm{C}Q  q_i \right., \label{22} \\
  S_{\mathrm{C}} &=& 0, \label{23} \\
  P_\mathrm{C}&=&-E_\mathrm{C}+ \sum_i \mu_{i,\mathrm{C}} n_i-V \frac{\partial E_\mathrm{C}}{\partial V} \nonumber \\
              &=& \frac{2}{9}\pi R^2 \alpha_\mathrm{C}
    ( Q^2 - Q_\mathrm{v}^2  )-\frac{4}{9}\pi R^3 \alpha_\mathrm{C}Q\sum_{j}q_{j}\frac{\partial n_{j}}{\partial R}.
\end{eqnarray}

The total pressure, entropy and chemical potential of a strangelet then becomes
\begin{eqnarray}
P_\mathrm{total} &=& P +P_\mathrm{C}, \label{16}\\
S_\mathrm{total} &=& S +S_\mathrm{C}, \label{17}\\
\mu_{i,\mathrm{total}} &=&\mu_{i}^{*}+\frac{1}{3}\frac{\partial m_{i}}{\partial n_\mathrm{b}}\frac{\partial\Omega_{0}}{\partial m_{i}}+\mu_{i,\mathrm{C}}. \label{18}
\end{eqnarray}

\section{Density and/or temperature dependent particle masses}
\label{sec:therm}

The equivalent quark masses are usually parametrized as
\begin{equation}
m_i=m_{i0}+m_\mathrm{I},
\end{equation}
where $m_\mathrm{I}$ accounts for the strong interaction and $m_{i0}$ the current quark mass with $m_{u0}=5\ \mathrm{MeV}$, $m_{d0}=10\ \mathrm{MeV}$, and $m_{s0}=100\ \mathrm{MeV}$.

According to MIT bag model, the quark mass scaling for equivparticle model was initially parameterized as~\cite{Fowler.9.271, Chakrabarty1989, Chakrabarty1991, Chakrabarty1993}
 \begin{equation}\label{27}
m_\mathrm{I}=\frac{B}{3n_\mathrm{b}},
\end{equation}
where $B$ is the bag constant. Based on the in-medium chiral condensates and linear confinement condition, a new quark mass scaling was later derived \cite{Peng_PhysRevC.61.015201}
\begin{equation}\label{28}
m_\mathrm{I}=\frac{D}{n_\mathrm{b}^{1/3}},
\end{equation}
where $D$ corresponds to the confinement parameter and is connected to the string tension, chiral symmetry restoration density and chiral condensate in vacuum. The one-gluon-exchange interaction can be considered as well by adding a new term \cite{Chen2012}, i.e.,
 \begin{equation}\label{29}
  m_\mathrm{I}= \frac{D}{n_\mathrm{b}^{1/3}}-C n_\mathrm{b}^{1/3},
 \end{equation}
where $C$ represents the strength of one-gluon-exchange interaction. By expanding $m_{I}$ as a Laurent series of Fermi momentum, the effects of perturbation interactions can be considered, where a new mass scaling was formulated as \cite{Xia_PhysRevD.89.105027}
\begin{equation}\label{30}
   m_\mathrm{I} = \frac{D}{n_\mathrm{b}^{1/3}}+C n_\mathrm{b}^{1/3}.
\end{equation}
Here $C$ takes positive values and represents the strength of first-order perturbation interaction. In order to describe the deconfinement phase transition, this formula is extended to the cases of finite temperature \cite{Lu2016}, i.e.,
\begin{eqnarray}\label{31}
m_\mathrm{I} &=& \frac{D}{n_\mathrm{b}^{1/3}}\left(1+\frac{8T}{\Lambda}e^{-{\Lambda}/{T}}\right)^{-1}\nonumber\\
             & & {} +Cn_{b}^{1/3}\left(1+\frac{8T}{\Lambda}e^{-{\Lambda}/{T}}\right),
\end{eqnarray}
where the constant $\Lambda = 280$ MeV.

In addition, we need to know the equivalent mass of gluons since their contributions can not be neglected. The gluon mass has been described according to a fast convergent expression of the QCD coupling base on the recent lattice data \cite{Borsanyi.Endrodi.ea1-22JHEP}. Borsanyi provided 48 values of the pressure by the lattice simulations, where we use the least square method to get the most effective fitting result. Here, we define the scaled temperature as $x= T/T_{c}$, where $T_{c}$ is the critical temperature. For $T<T_{c}$, the expression of gluon's equivalent mass is
\begin{equation}\label{32}
  \frac{m_{g}}{T}=\sum_{i}a_{i}x^{i}=a_{0}+a_{1}x+a_{2}x^{2}+a_{3}x^{3},
\end{equation}
with $a_{0}=67.018$, $a_{1}=-189.089$, $a_{2}=212.666$, $a_{3}=-83.605$. For $T> T_{c}$, the expression of gluon's equivalent mass is
\begin{equation}\label{33}
  \frac{m_{g}}{T}=\sum_{i}b_{i}\alpha^{i}=b_{0}+b_{1}\alpha+b_{2}\alpha^{2}+b_{3}\alpha^{3},
\end{equation}
where $b_{0}=0.218$, $b_{1}=3.734$, $b_{2}=-1.160$, $b_{3}=0.274$. The QCD coupling constant $\alpha=\alpha_{s}/\pi=g^{2}/(4\pi^{2})$ depends on the renormalization scheme and runs according to the renormalization equation. The expansion equation of $\alpha$ can be obtained by solving the renormalization group equation~\cite{06Peng413-418PLB}, which gives
\begin{eqnarray}\label{34}
\alpha=\frac{\beta_0}{\beta_0^2\ln(u/\Lambda)+\beta_1\ln\ln(u/\Lambda)}.
\end{eqnarray}
Here $\beta_0={11}/{2}-{N_f}/{3}$, $\beta_1={51}/{4}-{19N_f}/{12}$, ${u}/{\Lambda}=\sum_i c_ix^i=c_0+c_1x$, $c_{0}=1.054$, $c_{1}=0.479$.

\section{properties of strangelets }
\label{sec:SLT}

To study the properties of strangelets, the finite size effects need to be considered, which is treated with the multiple reflection expansion (MRE) method~\cite{Berger1987_PRC35-213, Madsen_PhysRevLett.70.391,Madsen_PhysRevD.47.5156,Madsen_PhysRevD.50.3328}. The MRE method introduces a modification to the density of states, which is given by
\begin{equation}\label{35}
   \rho_{i}(p,R,m_{i})=\frac{d_{i}p^{2}}{2\pi^{2}V}\Big[V+\frac{S_{r}}{p}f_{S}\big(\frac{m_{i}}{p}
   \big)+\frac{C_{r}}{p^{2}}f_{C}\big(\frac{m_{i}}{p}\big)\Big].
\end{equation}
The index $i$ represents the particle type, while $V$, $S_{r}$ and $C_{r}$ are the volume, surface area \cite{Berger1987_PRC35-213} and curvature term~\cite{Madsen_PhysRevLett.70.391, Madsen_PhysRevD.47.5156, Madsen_PhysRevD.50.3328} of a strangelet.  For a spherical system, $V=4\pi R^{3}/3$, $S_{r}=4\pi R^{2}$, $C_{r}=8\pi R$. The degeneracy factors are $d_{q}=3(colors)\times2(spins)=6$ for quarks ($q=u,d,s$) and $d_{g}=8(colors)\times2(spins)=16$ for gluons. The functions $f_{S}$ and $f_{C}$ are given as follows
 \begin{eqnarray}\label{36}
   f_{S}\big(\frac{m_{i}}{p}\big)&=&-\frac{1}{2}\mathrm{arctan}\big(\frac{m_{i}}{p}\big), \\
   f_{C}\big(\frac{m_{i}}{p}\big)&=&\frac{1}{6}\Big[1-\frac{3p}{2m_{i}}\mathrm{arctan}\big(\frac{m_{i}}{p}
   \big)\Big]. \label{37}
\end{eqnarray}
According to the equivparticle model, the thermodynamic potential density of free particles is
 \begin{equation}\label{38}
  \Omega_{0}=\Omega_{0}^{+}+\Omega_{0}^{-}+\Omega_{0}^{g}.
\end{equation}
The contributions of gluons $\Omega_{0}^{g}$, quarks $\Omega_{0}^{+}$ and antiquarks $\Omega_{0}^{-}$ are given by
\begin{eqnarray}
  \Omega_{0}^{g}   &=& T\int_{0}^{\infty}\mathrm{ln}\Big[1-e^{-\sqrt{p^{2}+m_{g}^{2}}/T}\Big]\rho_{g}\mathrm{d}p, \label{39} \\
  \Omega_{0}^{\pm} &=& -\sum_{i}\int_{0}^{\infty} T \ln\Big[1+e^{-(\sqrt{p_{}^{2}+m_{i}^2}\mp \mu_{i}^{*})/T}\Big] \rho_{i}\mathrm{d}p. \label{40}
\end{eqnarray}
Based on Eqs.~(\ref{16})-(\ref{18}), the specific expressions of each thermodynamic quantity at finite temperature can then bes obtained.

We know that there are weak interaction processes $d$,$ s  \leftrightarrow  u + e + \overline{\nu}_{e} $ and $s + u \leftrightarrow  u + d $ in strangelets. Because neutrino can freely enter and exit strange quark matter and the electron Compton wavelength is much larger than the size of strangelets with $A\ll 10^{7}$, the chemical potentials of neutrinos and electrons are thus zero, i.e.,
\begin{equation}\label{41}
\mu_{u,\mathrm{total}}=\mu_{d,\mathrm{total}}=\mu_{s,\mathrm{total}}
\end{equation}
with the chemical potential determined by Eq.~(\ref{18}). For a given baryon number $A$, the baryon number conservation needs to be fulfilled, i.e.,
\begin{equation}\label{43}
  A=\frac{1}{3}\sum_{i}N_{i}=\frac{1}{3}\sum_{i}n_{i}V.
\end{equation}
In addition, the pressure of the strangelets must be zero to maintain mechanically stablity, i.e.
\begin{equation}\label{44}
P=0.
\end{equation}

Therefore, we could obtain the energy density, particle number density, pressure, entropy and chemical potential by solving Eqs.~(\ref{41})-(\ref{44}). We define the ratios of charge number to baryon number and strangeness quantum number to quark number as
\begin{eqnarray}
f_z &=& Z/A = {\left(\frac{2}{3}N_u-\frac{1}{3}N_d-\frac{1}{3}N_s\right)}/{A}, \label{45} \\
f_s &=& S/\sum_{q}{N_{q}} = N_s/3A. \label{46}
\end{eqnarray}

\begin{figure}
\centering
\includegraphics[width=8cm]{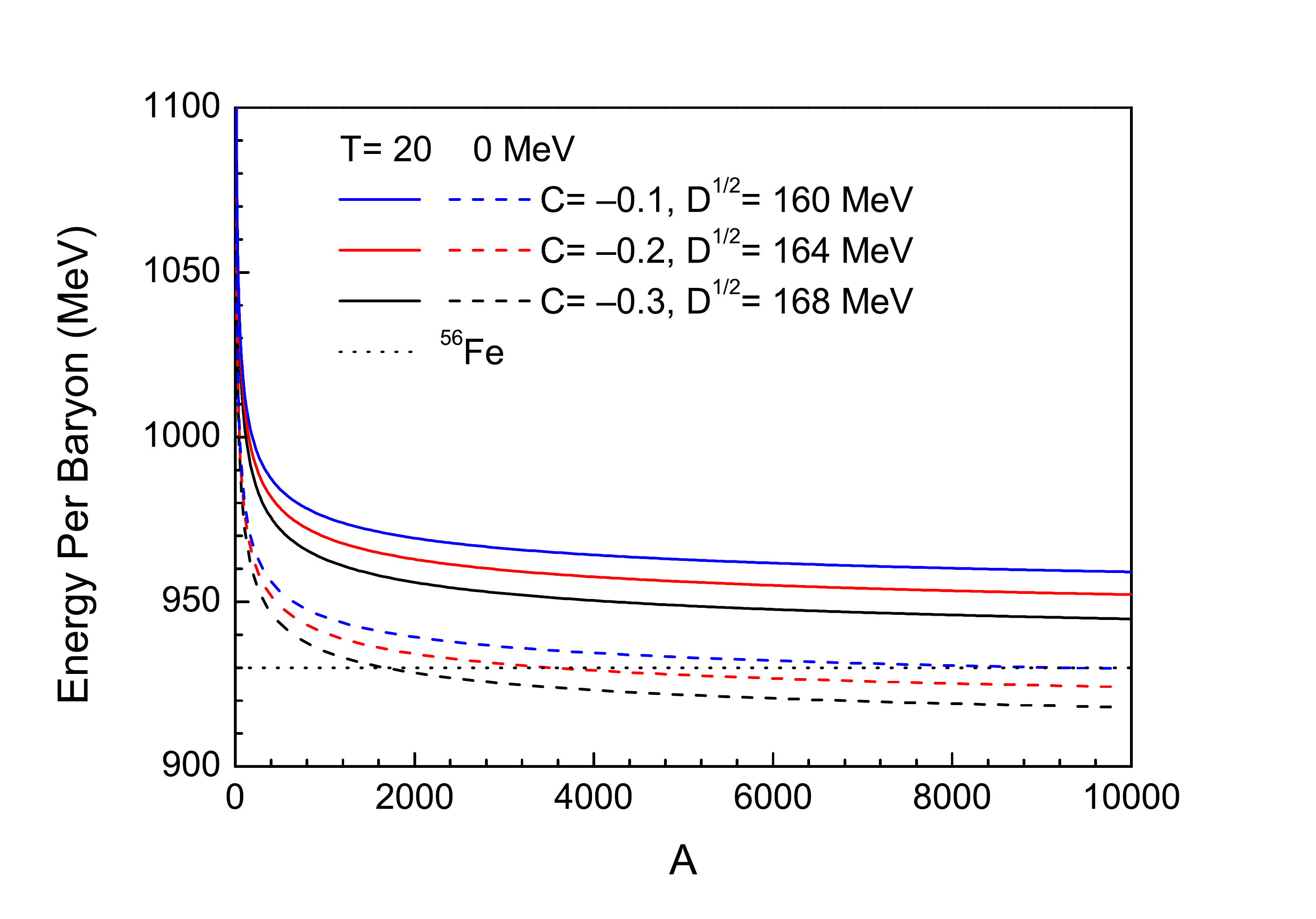}
\caption{Energy per baryon of $\beta$-stable strangelets as functions of baryon number.}
\label{Fig1}
\end{figure}

In Fig.~\ref{Fig1} we present the energy per baryon as a function of the baryon number $A$ at $T=0$ $\mathrm{MeV}$ and $T=20$ $\mathrm{MeV}$ for fixed parameters $C$ and $D$ \cite{Xia_PhysRevD.89.105027}. The solid, dashed and dotted curves correspond, respectively, to the temperature $T=20$, $0$ $\mathrm{MeV}$ and $^{56}Fe$. The blue, red and black curves correspond, respectively, to the parameter sets ($C$, $\sqrt{D}$ in MeV): $(-0.1,160)$, $(-0.2,164)$ and $(-0.3,168)$. It can be seen from the Fig.~\ref{Fig1} that the strangelets under three sets of parameters are absolutely stable at zero temperature, but not at $T=20$ $\mathrm{MeV}$. The energy per baryon of strangelets decreases with increasing baryon number $A$. It's worth mentioning that strangelets would emit neutrons when $E/A>m_{n} \approx 939$ $\mathrm{MeV}$ \cite{Berger1987_PRC35-213}. This suggests that strangelets with large baryon numbers could be absolutely stable at small temperature, which provide the possibility of detecting strangelets from the particles emitted by unknown astrophysical sources where those low temperature and high baryon numbers are expected.

\begin{figure}
\centering
\includegraphics[width=8cm]{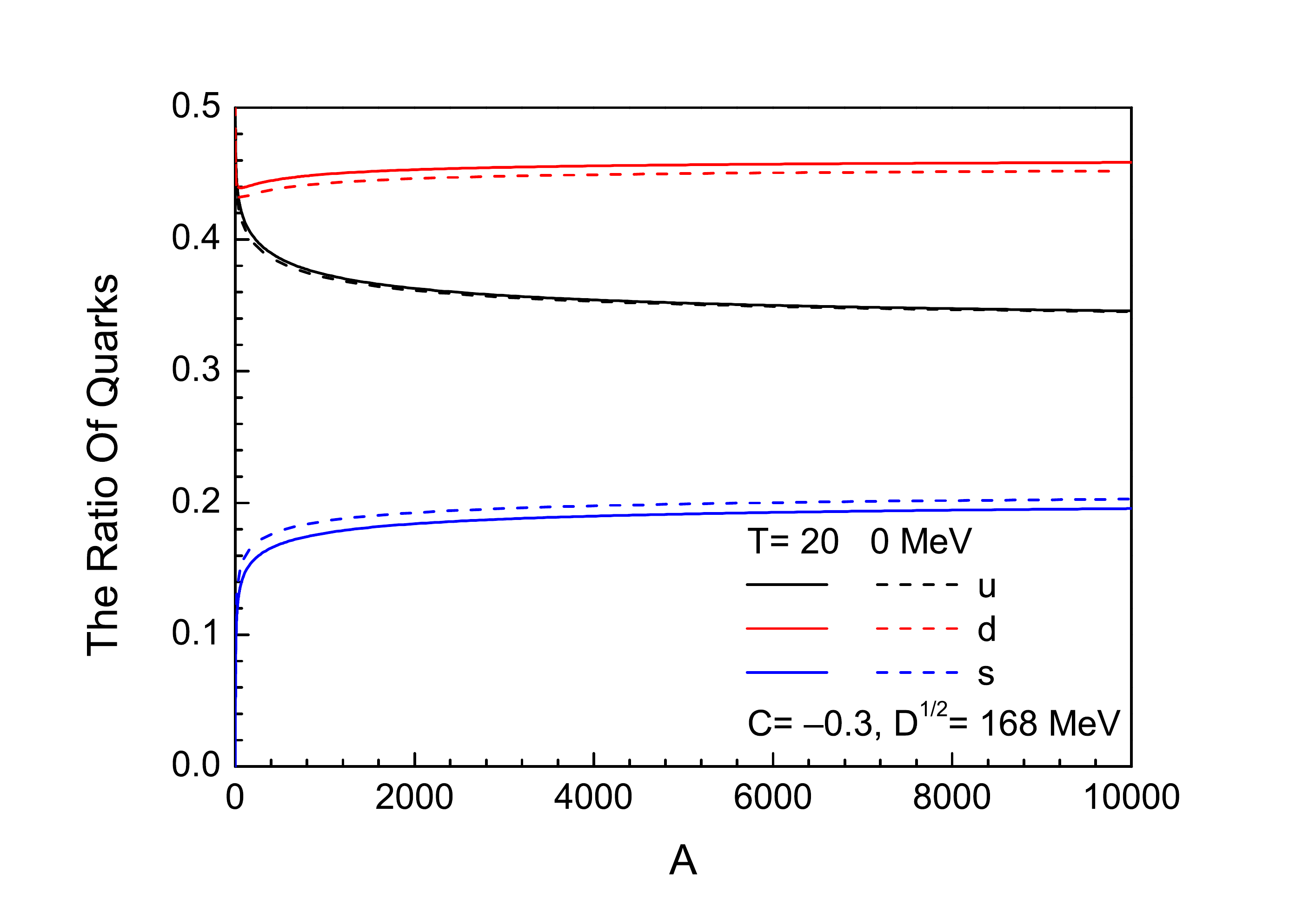}
\caption{The ratios of quarks to baryon number for $\beta$-stable strangelets as functions of baryon number.}
\label{Fig2}
\end{figure}

We can see the dependence of the proportion of the $u$, $d$ and $s$ quark on the baryon number from Fig.~\ref{Fig2}. The black, red and blue curves correspond, respectively, to the $u$, $d$ and $s$ quark. With the increasing of $A$, the proportion of $u$ quark is decreasing, while that of $d$ and $s$ quark increases.
That's because the number of weak interaction processes $d$,$ s  \leftrightarrow  u + e + \overline{\nu}_{e} $ and $s + u \leftrightarrow  u + d $ in the direction of $s$ increases accompany with the increase of $A$.

\begin{figure}
\centering
\includegraphics[width=8cm]{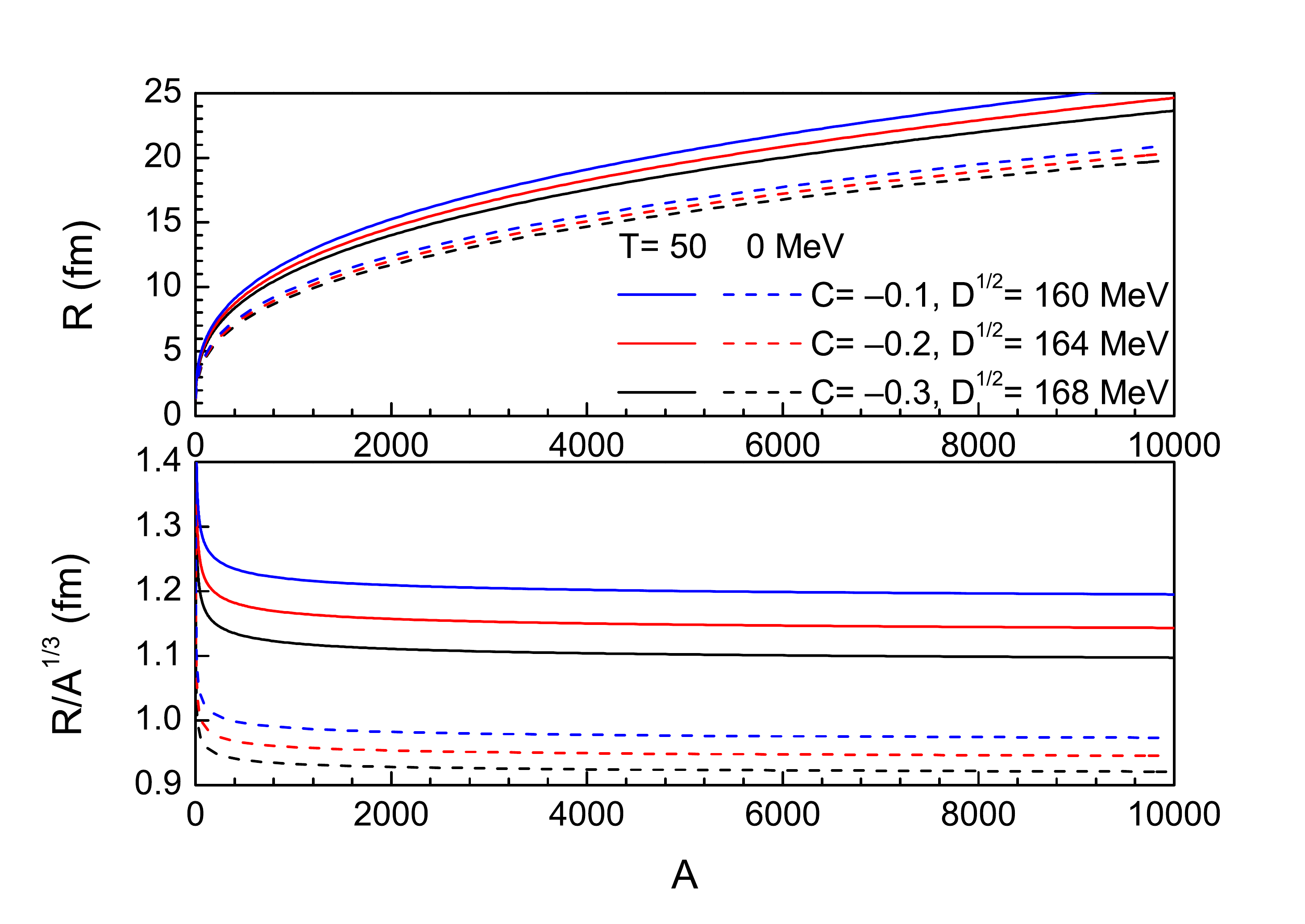}
\caption{Mechanically stable radii of $\beta$-stable strangelets as functions of baryon number.}
\label{Fig3}
\end{figure}

In Fig.~\ref{Fig3}, we show the dependence of mechanically stable radius of strangelets on the baryon number at $T=0$ and $T=20$ $\mathrm{MeV}$.
The solid and dashed curves correspond, respectively, to the temperature $T=50$ and $0$ $\mathrm{MeV}$. The blue, red and black curves correspond, respectively, to the parameter sets ($C$, $\sqrt{D}$ in MeV): $(-0.1,160)$, $(-0.2,164)$ and $(-0.3,168)$.
It can be seen from Fig.~\ref{Fig3} that the mechanically stable radius $R$ increases with baryon number $A$ while the ratio $R/A^{1/3}$ decreases. Moreover, the ratio $R/A^{1/3}$ decreases slowly and eventually tends to be a constant as $A\rightarrow \infty$, corresponding to a cube root relation $R=r_{0}A^{1/3}$ with a constant $r_{0}$.

\begin{figure}
\centering
\includegraphics[width=8cm]{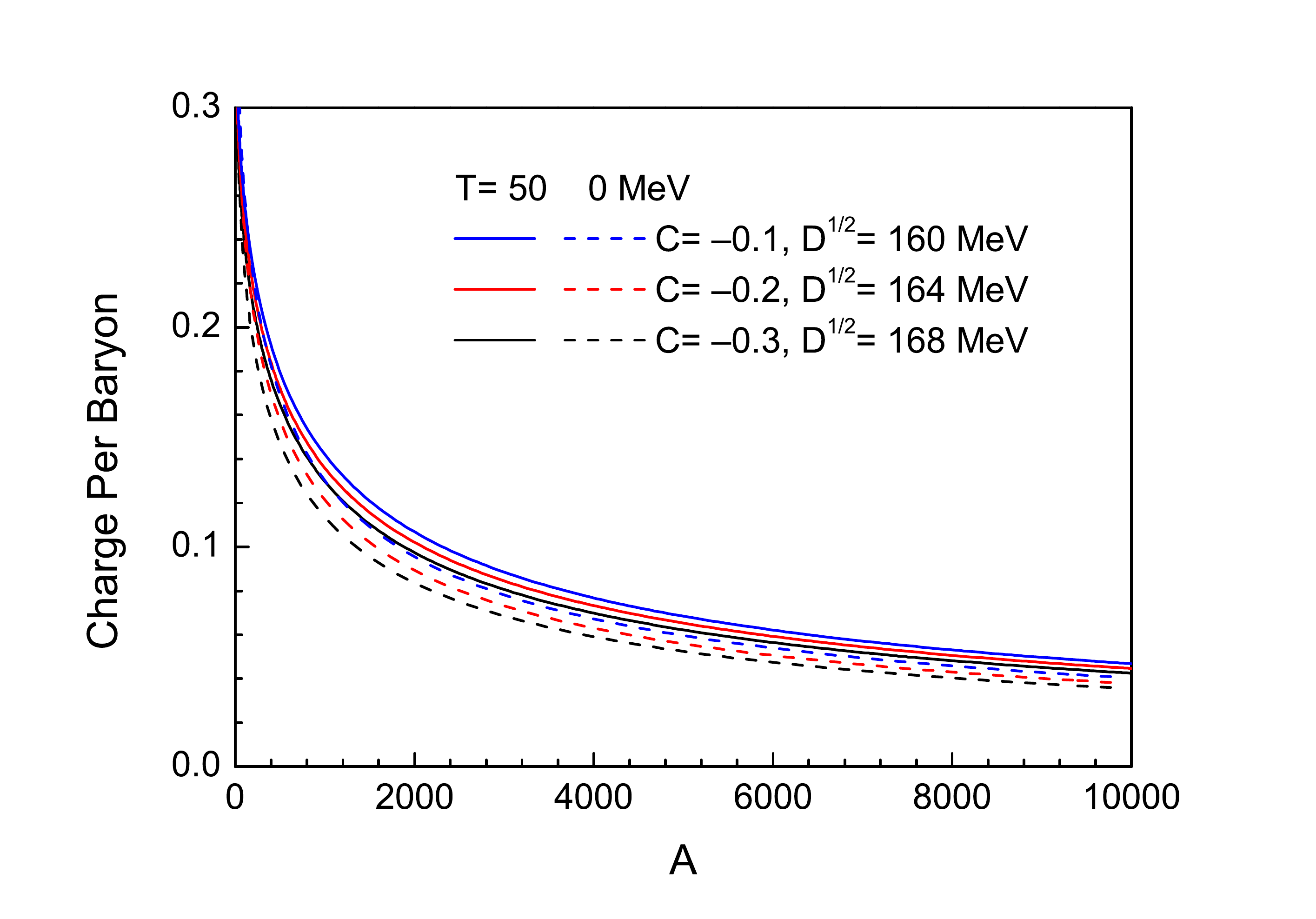}
\caption{The charge-to-mass ratio $f_z$ of $\beta$-stable strangelets as functions of baryon number.}
\label{Fig4}
\end{figure}

The dependence of the charge-to-mass ratio $f_z$ of the strangelets on the baryon number is depicted in Fig.~\ref{Fig4} at $T=0$ $\mathrm{MeV}$ and $T=50$ $\mathrm{MeV}$.
It can be seen that $f_z$ decreases with baryon number $A$ and tend to zero when the baryon number $A$ is large. This is consistent with the prediction of electrically neutral strange quark matter and the result in Fig.~\ref{Fig2}. Note that $f_z \rightarrow 0$ if we don't consider the contribution of Coulombic interactions. Furthermore, we notice that the one gluon exchange interaction reduce $f_z$.

\begin{figure}
\centering
\includegraphics[width=8cm]{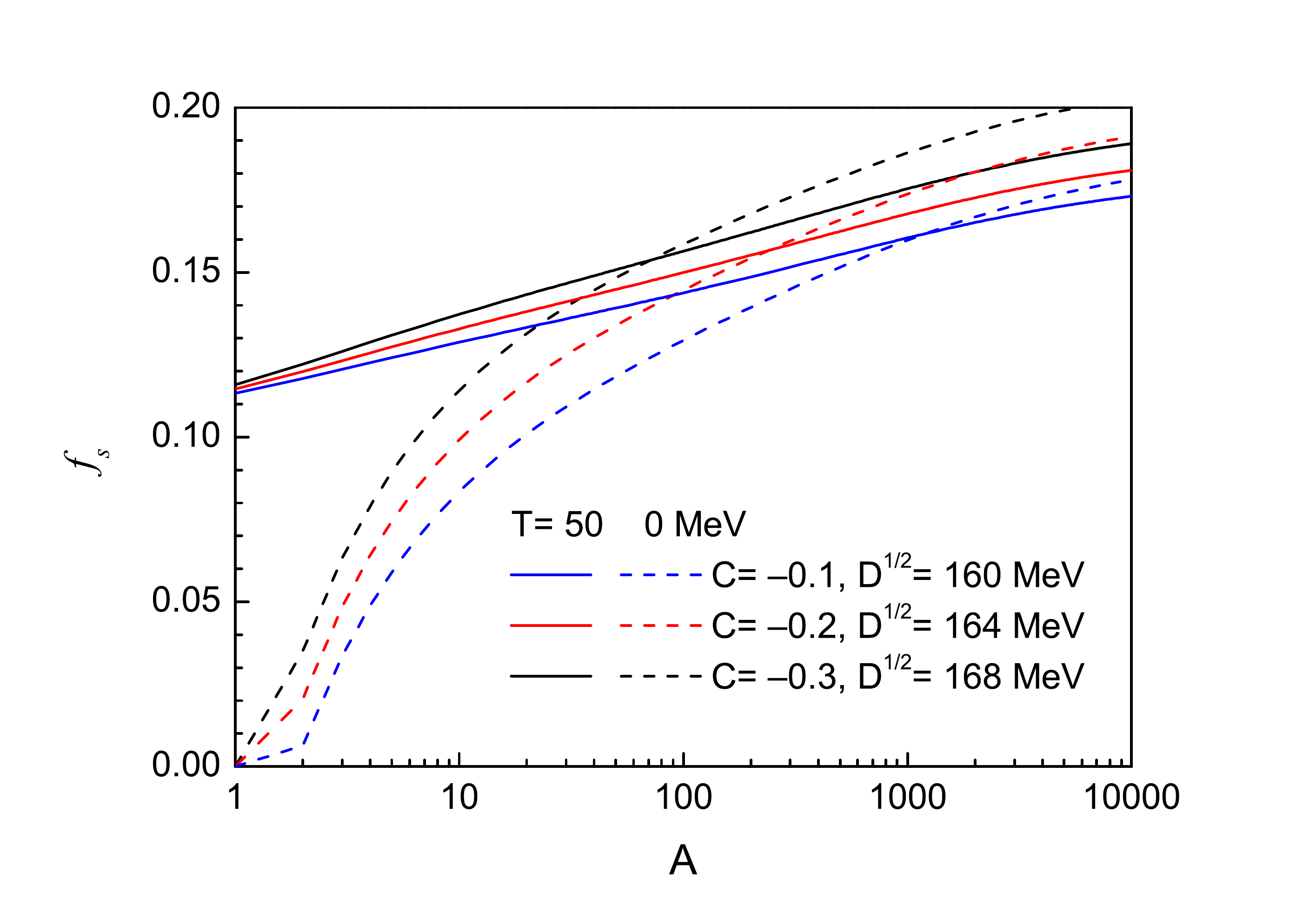}
\caption{The ratio of the strange quark number to total quark baryon number $f_{s}$ as functions of baryon number.}
\label{Fig5}
\end{figure}

\begin{figure}
\centering
\includegraphics[width=8cm]{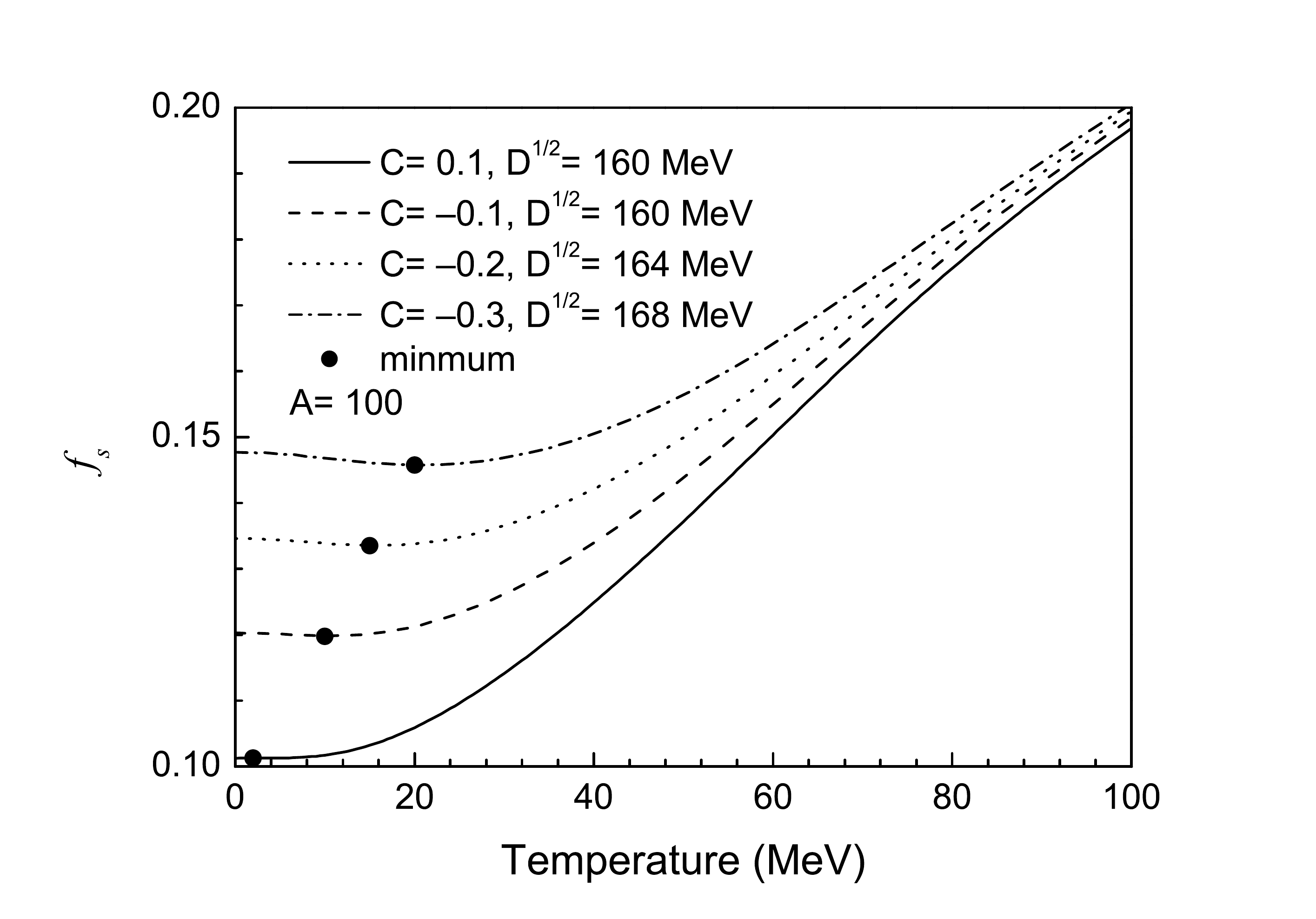}
\caption{The ratio of the strange quark number to total quark baryon number $f_{s}$ as functions of temperature when $A$ is 100.}
\label{Fig6}
\end{figure}

We use the ratio of the strange quark number to total quark numbe $f_s=N_s/3A$ to express the strangeness of strangelets, which is shown in Fig.~\ref{Fig5}. It can be seen $f_s$ is increasing and tends to a stable value when the baryon number $A$ is large. Furthermore, for strangelets with small $A$, we find that the strangeness gets larger for higher temperatures, while the trend is reversed at large $A$. The strangeness of strangelets tends to a constant value at any temperature. There are some interesting point, which we will explain it combination with Fig.~\ref{Fig6}.

In Fig.~\ref{Fig6}, we can see a minimum point of the strangeness of strangelets for fixed parameter $C$ and $D$ at $T=T_{0}$. The solid, dashed, dotted and dash dotted curves correspond, respectively, to the parameter sets $(C,\sqrt{D})$: $(0.1,160)$, $(-0.1,160)$, $(-0.2,164)$ and $(-0.3,168)$. It is found that the one gluon exchange interaction increases $f_s$ and $T_{0}$, while the first-order perturbation interaction reduces $f_s$ and eventually $T_{0}$ reaches zero.

\begin{figure}
\centering
\includegraphics[width=8cm]{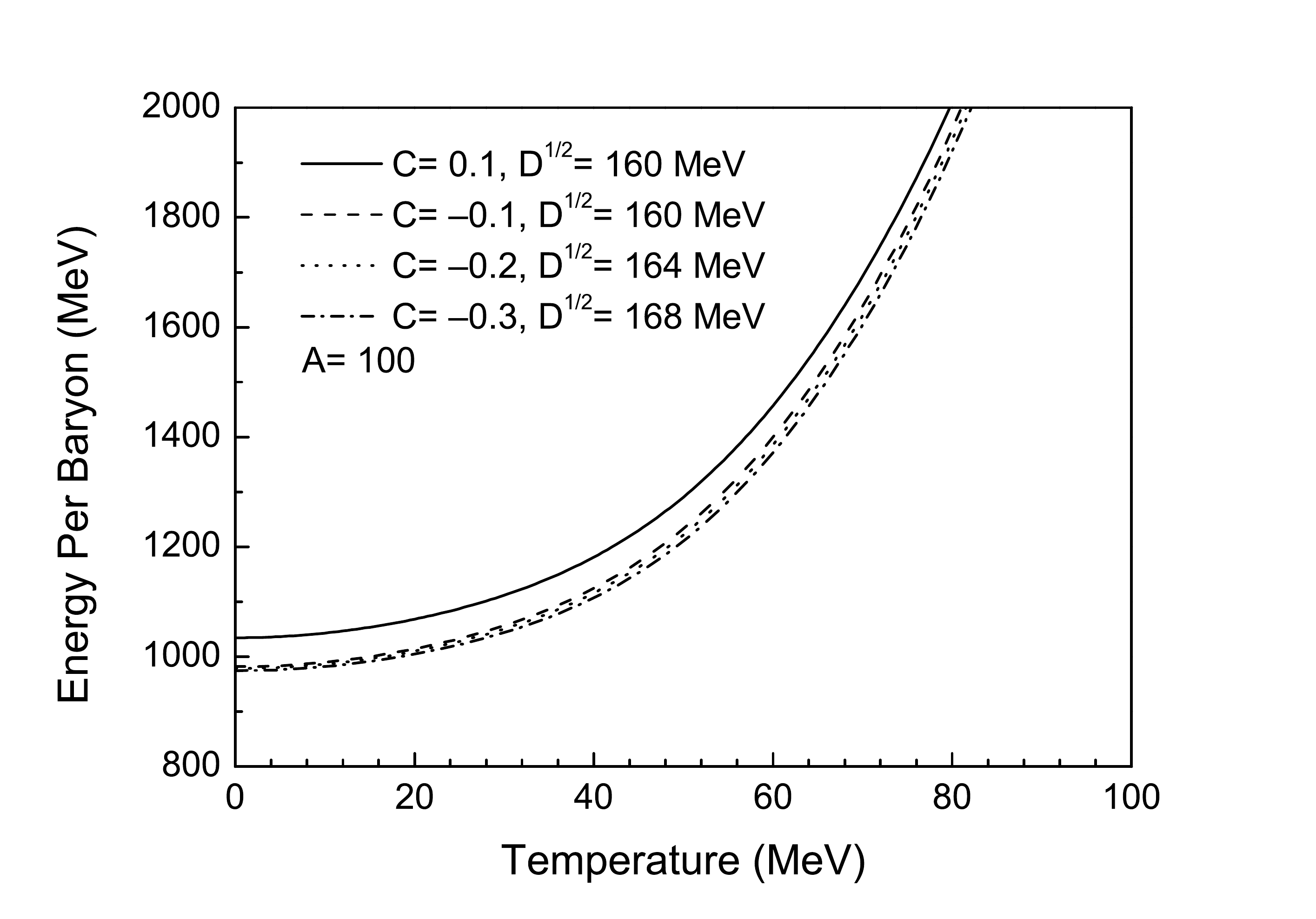}
\caption{The energy per baryon as functions of the temperature when $A$ is 100.}
\label{Fig7}
\end{figure}

Next, we study the relationship between the temperature and energy per baryon in Fig.~\ref{Fig7}. We choose the parameter sets $(C,\sqrt{D})$: $(0.1,160)$, $(-0.1,160)$, $(-0.2,164)$ and $(-0.3,168)$. We can see that the higher the temperature, the great the energy per baryon is at $A=100$ due to the increment of the thermal energy of quarks and gluons. Therefore, strangelets will become unstable and are not likely to exist. In addition, it is found that the one gluon exchange interaction reduces the energy and the perturbative interaction increases the energy.

\begin{figure}
\centering
\includegraphics[width=8cm]{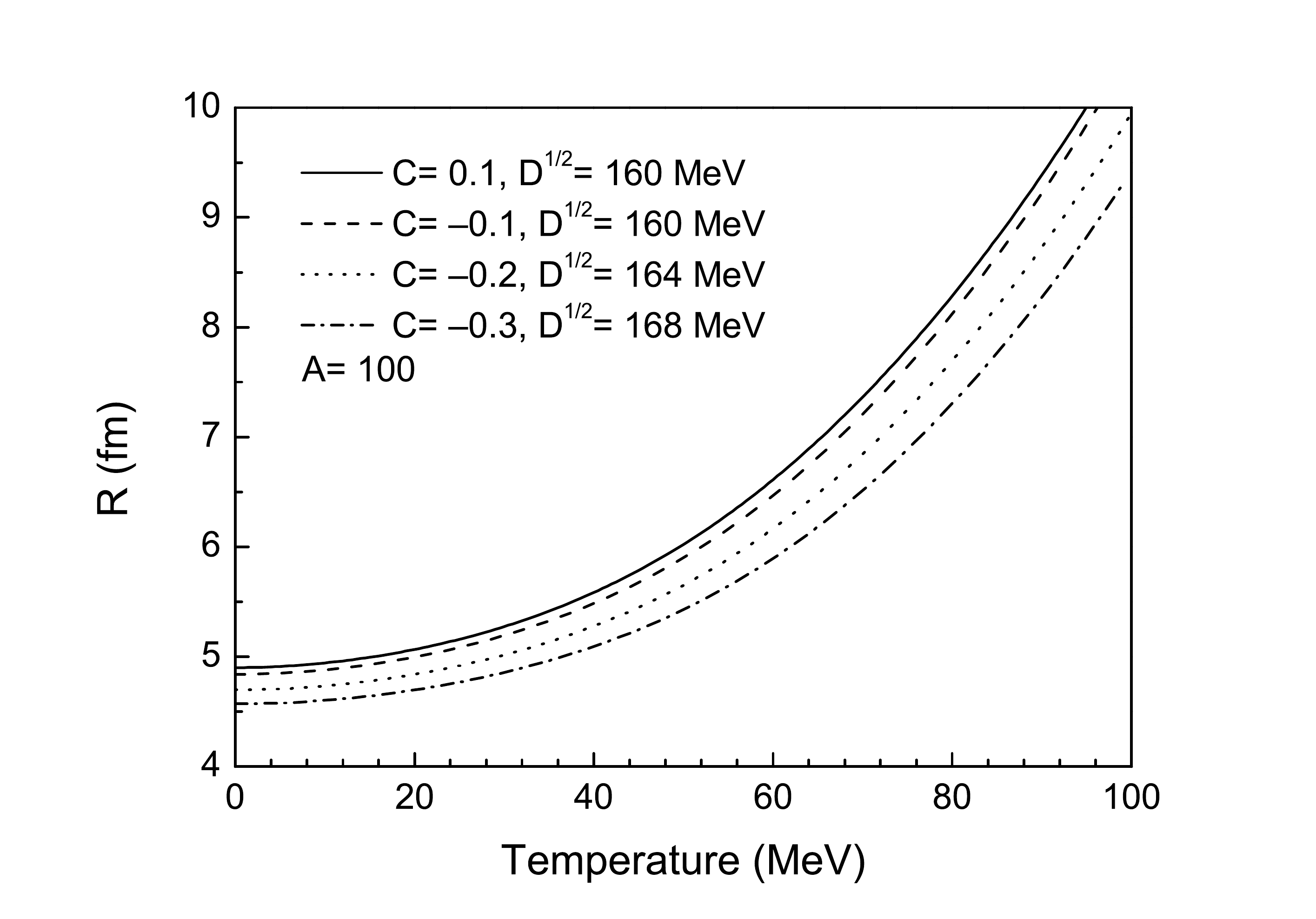}
\caption{The mechanically stable radius as functions of the temperature when $A$ is 100.}
\label{Fig8}
\end{figure}

In Fig.~\ref{Fig8}, we note that the mechanically stable radius is a monotone increasing function of the temperature when $A$ is 100. Furthermore, we notice that there should be an exponential function relation between $R$ and $T$. This is consistent with the result of the energy per baryon in Fig.~\ref{Fig7}. We also find the perturbative interaction increases the mechanically stable radius while the one-gluon-exchange interaction decreases the mechanically stable radius at fixed $D$.

\begin{figure}
\centering
\includegraphics[width=8cm]{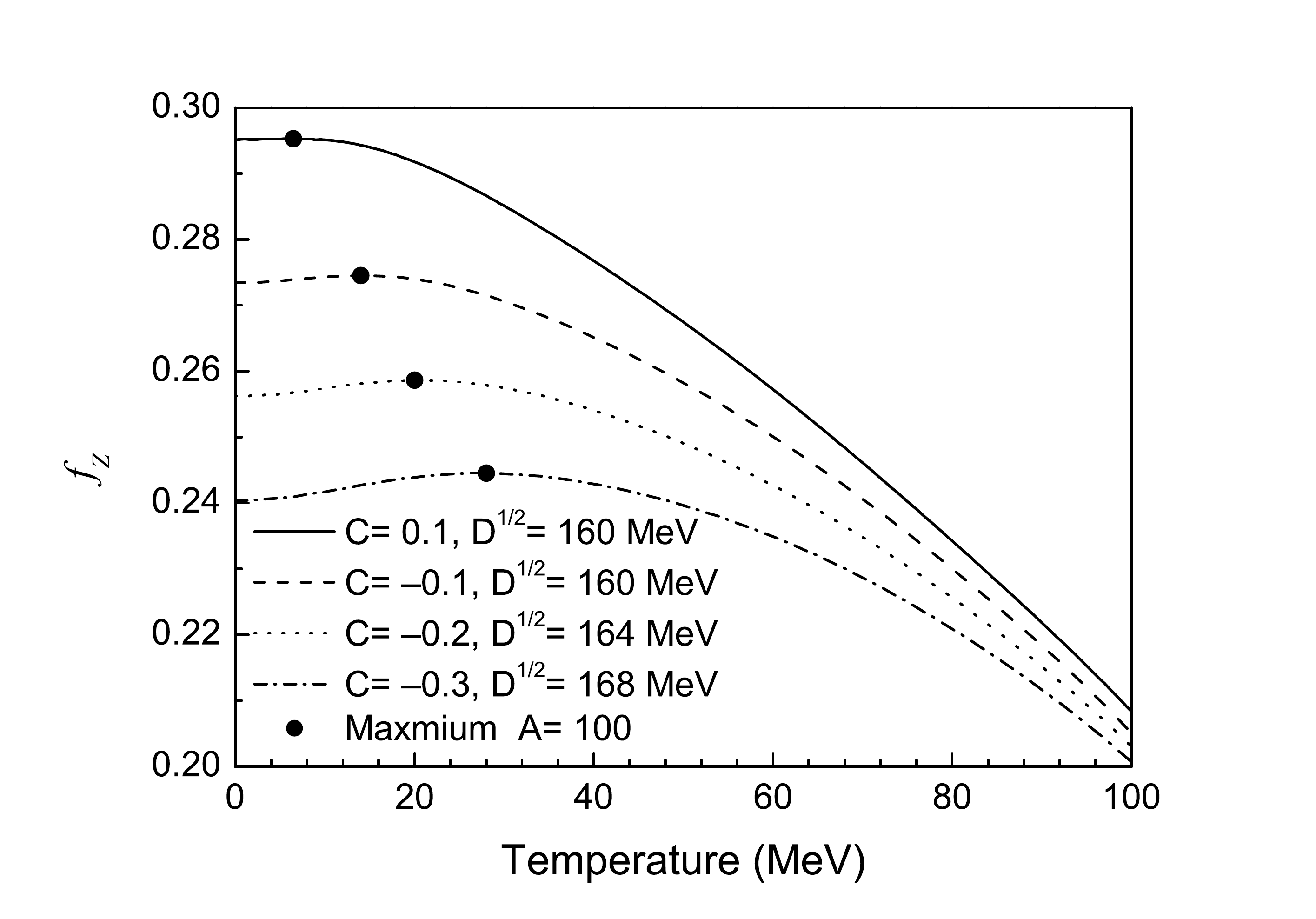}
\caption{The ratio of charge number to the baryon number $f_z$ as functions of the temperature when $A$ is 100.}
\label{Fig9}
\end{figure}

In Fig.~\ref{Fig9}, we present the charge to mass ration as a function of temperature when $A$ is 100. We see that $f_{z}$ is not just negativly correlate with temperature, where a maximum point exists for the strangeness of strangelets. Let's say the temperature at this point is $T_{\mathrm{max}}$. At $T<T_{\mathrm{max}}$, the higher the temperature, the larger the strangeness of strangelets, while otherwise the opposite at $T>T_{\mathrm{max}}$. Note that the different parameter sets $(C,\sqrt{D})$ have different maximum point. Furthermore, we see that the first-order perturbative interaction increases $f_z$ and reduces $T_{\mathrm{max}}$, and the one-gluon-exchange interaction decreases $f_z$ and increases $T_{\mathrm{max}}$.
This is consistent with the result of $f_{s}$ in Fig.~\ref{Fig6}.

\begin{figure}
\centering
\includegraphics[width=8cm]{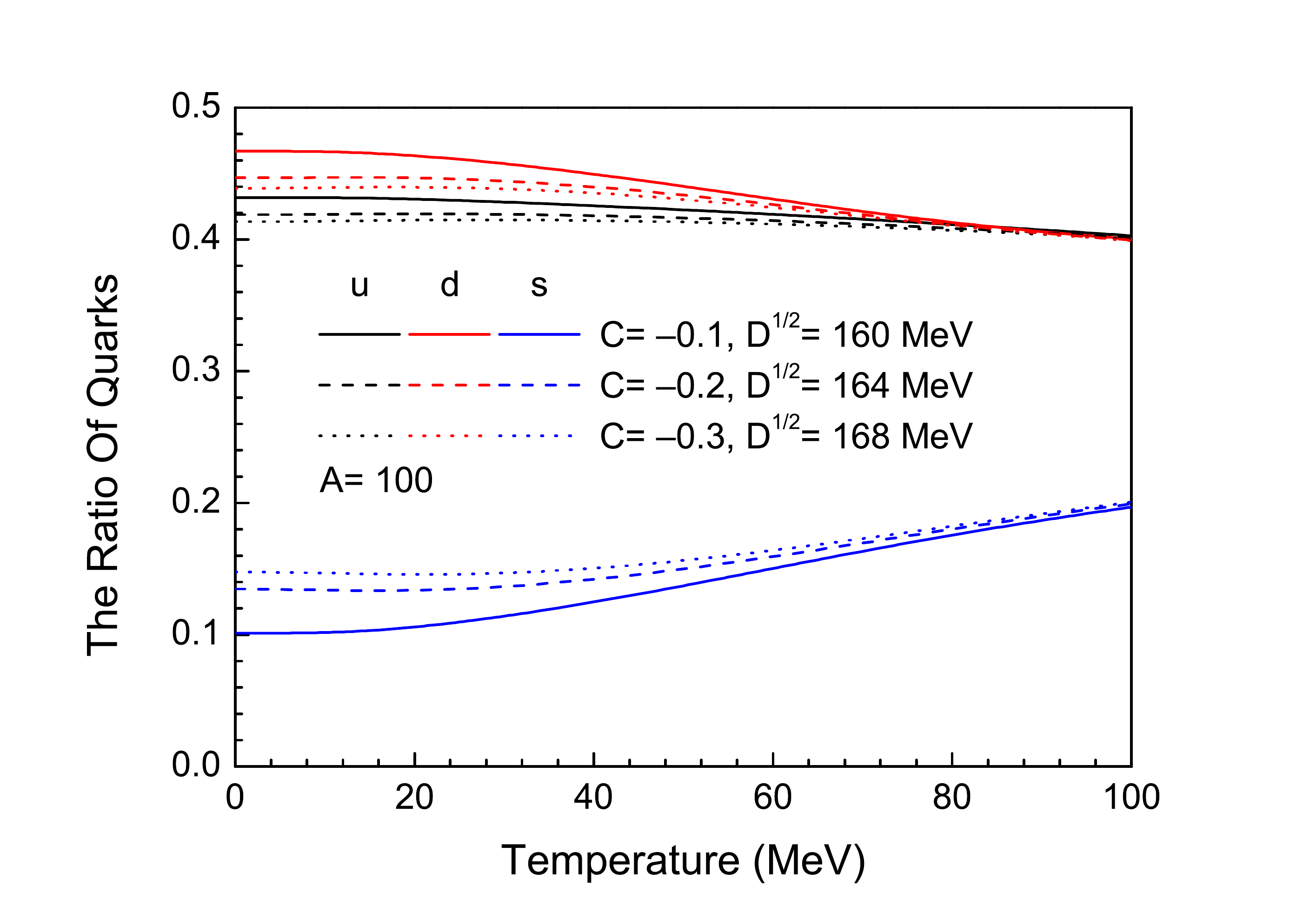}
\caption{The proportion of quarks as functions of the temperature when $A$ is 100.}
\label{Fig10}
\end{figure}

The dependence of the proportion of quarks on the temperature is depicted in Fig.~\ref{Fig10}.
The black, red and blue curves correspond, respectively, to the quark $u$, $d$ and $s$.
The solid, dashed, dotted and dash dotted curves correspond, respectively, to the parameter sets $(C,\sqrt{D})$: $(0.1,160)$, $(-0.1,160)$, $(-0.2,164)$ and $(-0.3,168)$.
With the increase of temperature, the proportion of $u$ and $d$ quark is decreasing, while that of $s$ quark increases. This is because $u$ and $d$ quark are converted to $s$ quark at high temperature. We also find that the first-order perturbation interaction increases the proportion of $u$ and $d$ quark and the one gluon exchange interaction decreases the proportion of $u$ and $d$ quark. At the same time, the first-order perturbation interaction decreases the proportion of $s$ quarks and the one-gluon-exchange interaction increases the proportion of $s$ quark.

\begin{figure}
\centering
\includegraphics[width=8cm]{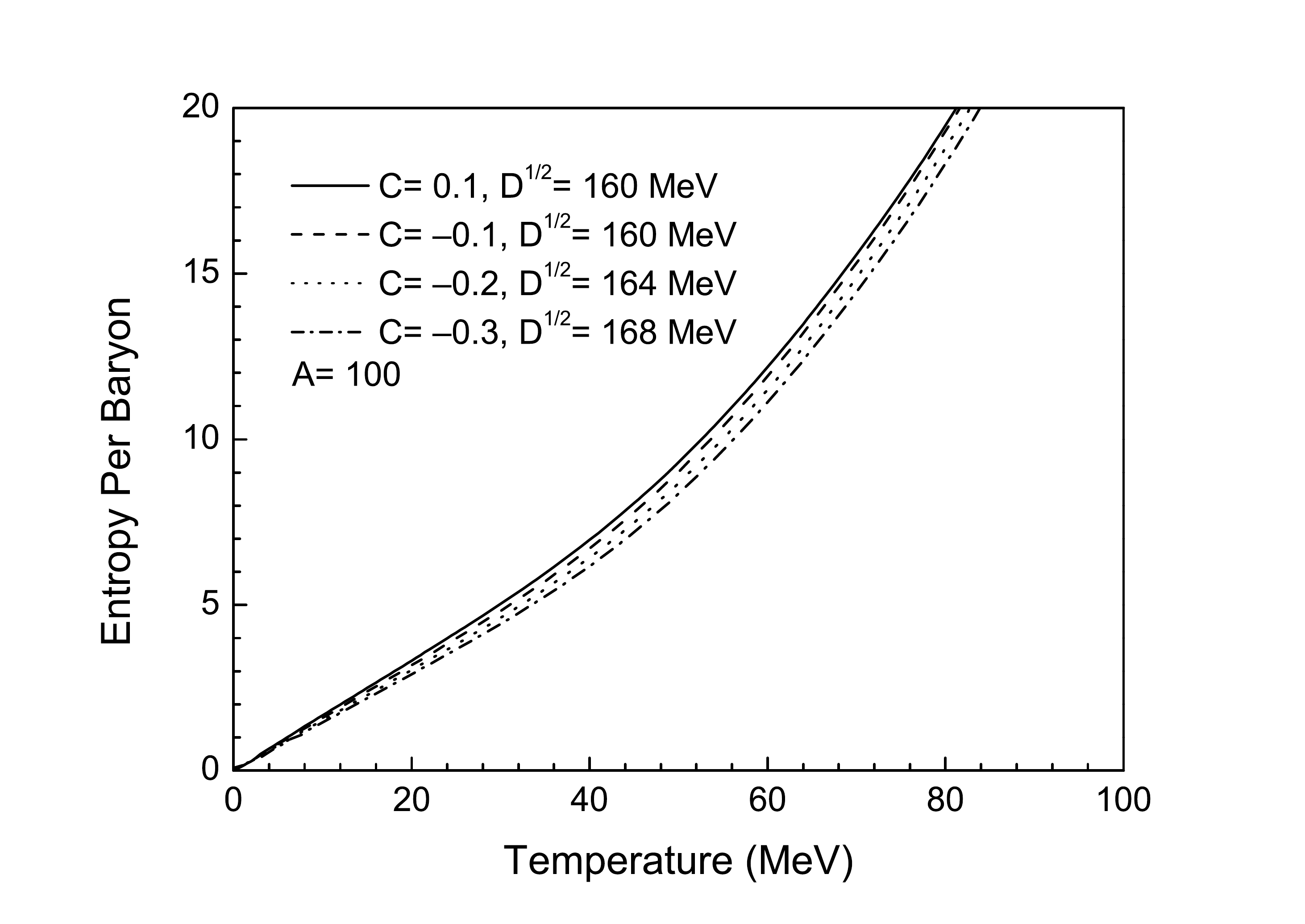}
\caption{The entropy per baryon as functions of the temperature for the fixed parameter sets $(C,\sqrt{D})$ when $A$ is 100.}
\label{Fig11}
\end{figure}

Figure~\ref{Fig11} shows the entropy per baryon as functions of temperature for the different parameter sets $(C,\sqrt{D})$.
The first-order perturbation interaction improves the entropy and the one gluon exchange interaction reduces the entropy.
It's an increasing function of temperature, and approaches to zero at vanishing temperature.
This is guaranteed by
\begin{eqnarray}
\lim_{T\rightarrow 0}\partial m_{q} /\partial T=0.
\end{eqnarray}

\section{SUMMARY}
\label{sec:sum}

We have investigated the properties of strangelets at finite temperature by the equivparticle model with a quark mass scaling with  confinement and perturbative interaction. We have analyzed the contribution of Coulomb interaction by a thermodynamic self-consistent way and obtained the contributions to the pressure and chemical potential. Combine the equivparticle model with self-consistent thermodynamic treatment, we studied the properties of strangelets, such as the energy per baryon of strangelets, the mechanically stable radius of strangelets, strangeness, charge property and the proportion of quarks. It is found that the energy per baryon and $f_z$ decreased and the mechanically stable radius and strangeness increased as baryon number $A$ increases. In addition, the obtained energy per baryon, mechanically stable radius and strangeness increased with temperature and $f_z$ decreased.

\appendix

\section*{ACKNOWLEDGMENTS}

The authors would like to thank the support from National Natural Science Foundation
of China (Grant No. 11875052).

\end{document}